\documentclass[12pt]{article}
\usepackage{amsmath}
\usepackage[dvips]{graphicx}
\usepackage{epsfig}
\usepackage{amsmath}
\usepackage{amssymb}
\usepackage{graphics,epstopdf}
\usepackage{amsfonts}
\usepackage{amsthm}
\usepackage[usenames]{color}

\definecolor{AV}{rgb}{0.65,0.0,0}
\definecolor{GC}{rgb}{0,0.0,0.65}
\definecolor{WS}{rgb}{0,0.65,0}

\usepackage[
      colorlinks=true,
      linkcolor=blue,
      urlcolor=blue,
      filecolor=blue,
      citecolor=blue,
      pdfstartview=FitV,
      pdftitle={},
      pdfauthor={},
      pdfsubject={},
      pdfkeywords={},
      pdfpagemode=None,
      bookmarksopen=true
]{hyperref}

\newcommand{\bm}{\begin{multiline}}
\newcommand{\beq}{\begin{equation}}
\newcommand{\eeq}{\end{equation}}
\newcommand{\beqs}{\begin{eqnarray}}
\newcommand{\eeqs}{\end{eqnarray}}

\newcommand{\ra}{\rightarrow}

\begin{document}

\thispagestyle{empty}

\hfill{}

\hfill{}

\hfill{}

\vspace{32pt}

\begin{center}

\textbf{\Large  Thermodynamics of non-extremal Kaluza-Klein multi-black holes in five dimensions }

\vspace{48pt}

\textbf{ Cristian Stelea,}\footnote{E-mail: \texttt{cristian.stelea@uaic.ro.}}
\textbf{Kristin Schleich,$^2$~}
{\bf and Donald Witt$^2$}

\vspace*{0.2cm}

\textit{$^1$Faculty of Physics, ``Alexandru Ioan Cuza" University}\\[0pt]
\textit{11 Blvd. Carol I, Iasi, 700506, Romania}\\[.5em]

\vspace*{0.2cm}

{\it $^2$ Department of Physics and Astronomy, University of British Columbia}\\
{\it 6224 Agricultural Road, Vancouver, BC V6T 1Z1, Canada}\\[.5em]

\end{center}

\vspace{30pt}

\begin{abstract}

Using a solution-generating method, we derive an exact solution of the Einstein's field equations in five dimensions describing multi-black hole configurations. More specifically, this solution describes systems of non-extremal static black holes with Kaluza-Klein asymptotics. 
As expected, we find that, in general, there are conical singularities in-between the Kaluza-Klein black holes that cannot be completely eliminated. Notwithstanding the presence of these conical singularities, such solutions still exhibit interesting thermodynamical properties. By choosing an appropriate set of thermodynamic variables we show that the entropy of these objects still obeys the Bekenstein-Hawking law for spaces with Kaluza-Klein asymptotics. This extends the previously known thermodynamic description of asymptotically flat spaces with conical singularities to general spaces with Kaluza-Klein asymptotics with conical singularities. Finally, we obtain a charged generalization of this multi-black hole solution in the general Einstein-Maxwell-Dilaton theory and show how to recover the extremal multi-black hole solution as a particular case. 
 
\end{abstract}

\vspace{32pt}

\setcounter{footnote}{0}

\newpage

\section{Introduction}

Black holes in higher dimensions have been actively studied in recent years. Notably, with Emparan and Reall's discovery of the five dimensional black ring solution \cite{Emparan:2001wn}, it was realized that higher dimensional black holes exhibit a much richer structure than their four-dimensional counterparts (for nice reviews of the black ring see  \cite{Emparan:2006mm} and of higher dimensional black holes see \cite{Emparan:2008eg,Obers:2008pj}). In four-dimensional asymptotically flat space-times, as shown geometrically by Hawking \cite{Hawking:1971vc}, a black hole can have only spherical horizon topology; this result also follows under more general conditions as a consequence of topological censorship \cite{Chrusciel:1994tr,Jacobson:1994hs}. However, in five dimensions, the spherical topology of infinity does not constrain that of the black hole horizon \cite{Galloway:1999bp,Galloway:1999br}. Geometric considerations, however, restrict the horizon topology to those, such as $S^3$ and $S^2\times S^1$, that admit positive scalar curvature \cite{Galloway:2005mf}. In this regard, the black ring provides us with an explicit example of an asymptotically flat black hole having an $S^1\times S^2$ horizon topology. Furthermore, it  can carry (in certain conditions) the same amount of mass and angular momenta as the spherical Myers-Perry black hole \cite{Myers:1986un}. Consequently, five-dimensional black holes are not uniquely characterized by their mass and angular momenta; the uniqueness theorems for black holes in four dimensions cannot be extended to the five-dimensional case without further assumptions of additional symmetry and specification of the rod structure \cite{Hollands:2007aj}.

In five dimensions, there also exist the so-called squashed Kaluza-Klein (KK) black holes, whose horizon geometry is a squashed three-sphere \cite{Dobiasch:1981vh,Gibbons:1985ac,Rasheed:1995zv,Larsen:1999pp}. Their geometry is asymptotic to a non-trivial $S^1$ bundle over a four-dimensional asymptotically flat spacetime, which is also the asymptotic geometry of the Kaluza-Klein magnetic monopole \cite{Sorkin:1983ns,Gross:1983hb}. Such black holes look five-dimensional in the near-horizon region, while at infinity (asymptotically) they look like four-dimensional objects with a compactified fifth dimension. Again, uniqueness theorems for KK black holes are proven assuming additional symmetry and specification of other invariants \cite{Hollands:2008fm}. Thus explicit examples of such solutions are valuable. KK black hole solutions  in the presence of matter fields are generally found by solving the Einstein equations by brute force. For instance, a solution describing a static KK black hole with electric charge has been found in \cite{Ishihara:2005dp}, and the corresponding Einstein-Yang-Mills solution has been described in \cite{Brihaye:2006ws}. Remarkably, with hindsight, many  such KK solutions can be generated from known solutions by applying a `squashing' transformation on suitable geometries \cite{Wang:2006nw,Nakagawa:2008rm,Tomizawa:2008hw,Matsuno:2008fn,Tomizawa:2008rh,Stelea:2008tt}. However, not all of the KK black hole solutions can be generated by a squashing transformation; more general KK black holes have been derived  in the context of the minimal five-dimensional supergravity  \cite{Tomizawa:2008qr,Gal'tsov:2008sh,Mizoguchi:2011zj}.

In our work, we focus on multi-black hole solutions in spaces with Kaluza-Klein asymptotics. In general, in higher dimensions, solutions describing general systems of charged multi-black holes are rare. Unlike the single black hole case, all known solution-generating techniques lead to solutions describing configurations formed either from extremal black holes \cite{Myers:1986rx,Duff:1993ye,Ishihara:2006iv,Elvang:2005sa,Matsuno:2012hf} or from non-extremal black holes with charges proportional to the masses. In five dimensions, a solution describing a general double-black hole configuration has been recently constructed in \cite{Chng:2008sr,Stelea:2011jm}, generalizing the asymptotically flat solutions given in \cite{Tan:2003jz}. The solution generating technique from \cite{Chng:2008sr} has been further modified in  \cite{Stelea:2009ur} to obtain multi-black hole solutions in spaces with KK asymptotics. More precisely, a solution describing a system of two general KK black holes in the double-Taub-NUT background has been explicitly constructed and studied in  \cite{Stelea:2009ur}. One purpose of the present work is to construct explicitly a general exact solution describing a superposition of $N$ static Kaluza-Klein black holes with squashed horizons in five dimensions. As long as we are aware, this solution is unknown in literature. For simplicity purposes, we first consider the particular case of $N$ uncharged black holes. This solution is the five-dimensional Kaluza-Klein analog of the Israel-Khan solution, which describes a four-dimensional system of $N$ collinear black holes. A similar multi-black hole solution in five-dimensional asymptotically flat spaces has been previously studied in \cite{Tan:2003jz}. As generally expected on physical grounds, since these multi-black hole configurations are static, the presence of the conical singularities in between the black holes is unavoidable since they are needed to provide the necessary forces to balance the gravitational attraction among the black holes and keep the black hole system in equilibrium. For asymptotically flat spaces it turns out that even in the presence of conical singularities such singular geometries still admit a  reasonable thermodynamic description as recently shown in \cite{Costa:2000kf,Herdeiro:2010aq,Herdeiro:2009vd}. In this work we generalize that thermodynamic description to spaces with KK asymptotics in presence of conical singularities. Finally, by using a standard charging technique, we obtain the generalization to a solution describing a configuration of static electrically charged squashed black holes, with fixed mass-to-charge ratios. We show how to obtain the extremal version, obtaining this way a general extremal multi-black hole solution of the full EMD theory in five dimensions, generalizing the previously known solutions. We also comment on how to obtain the most general non-extremal solution of this kind in five dimensions. 

\section{The solution generating technique}

Let us recall here the main results of the solution generating technique used in \cite{Stelea:2009ur}. The main idea of this method is to map a general static electrically charged axisymmetric solution of Einstein-Maxwell theory in four dimensions to a five-dimensional static electrically charged axisymmetric solution of the Einstein-Maxwell-Dilaton (EMD) theory with arbitrary coupling of the dilaton to the electromagnetic field. To this end one performs first a dimensional reduction of both theories down to three dimensions and, after a careful comparison of the dimensionally-reduced lagrangians and mapping of the scalar fields and electromagnetic potentials, one is able to bypass the actual solving of the field equations by algebraically mapping solutions of one theory to the other. More precisely, suppose that we are given a static electrically charged solution of the four-dimensional Einstein-Maxwell system with Lagrangian
\begin{eqnarray}  \label{4delectric}
\mathcal{L}_4&=&\sqrt{-g}\left[R-\frac{1}{4}\tilde{F}_{(2)}^2\right],
\end{eqnarray}
where $\tilde{F}_{(2)}=d\tilde{A}_{(1)}$ and the only non-zero component of $\tilde{A}_{(1)}$ is $\tilde{A}_{t}=\Psi$. The solution to the equations of motion derived from (\ref{4delectric}) is assumed to have the following static and axisymmetric form:
\begin{eqnarray}  \label{4dKhan}
ds_{4}^{2} &=&-\tilde{f}dt^{2}+\tilde{f}^{-1}\big[e^{2\tilde{\mu}}(d\rho
^{2}+dz^{2})+\rho ^{2}d\varphi ^{2}\big],~~~~~~~
{\tilde{A}_{(1)}} =\Psi dt.
\end{eqnarray}
Here and in what follows we assume that all the functions $\tilde{f}$, $\tilde{\mu}$, and $\Psi$ depend only on coordinates $\rho$ and $z$.

Then the corresponding solution of the Einstein-Maxwell-Dilaton system in five dimensions with Lagrangian
\begin{eqnarray}
\mathcal{L}_{5}=\sqrt{-g}\left[R-\frac{1}{2}(\partial\phi)^2 -\frac{1}{4}e^{\alpha\phi}F_{(2)}^2\right]
\label{EMDaction5d}
\end{eqnarray}
where $F_{(2)}=dA_{(1)}$ can be written as:
\beqs
\label{final5dalpha}
ds_{5}^{2}&=&-\tilde{f}^{\frac{4}{3\alpha^2+4}}dt^{2}+\tilde{f}^{-\frac{2}{3\alpha^2+4}}\bigg[\frac{e^{2h}}{a^2-c^2e^{4h}}(d\chi+4acH d\varphi)^{2}+(a^2-c^2e^{4h})e^{\frac{6\tilde{\mu}}{3\alpha^2+4}
+2\gamma-2h}(d\rho ^{2}+dz^{2})\nonumber\\
&&+\rho^2(a^2-c^2e^{4h})e^{-2h}d\varphi ^{2}\bigg],~~~~~~~
A_{(1)}=\sqrt{\frac{3}{3\alpha^2+4}}\Psi dt,~~~~~~~ e^{-\phi}=\tilde{f}^{\frac{3\alpha}{3\alpha^2+4}}.
\eeqs
Here $a$ and $c$ are constants, while $h$ is an arbitrary harmonic function\footnote{That is, it satisfies the equation $\nabla^2h=\frac{\partial^2h}{\partial\rho^2}+\frac{1}{\rho}\frac{\partial h}{\partial\rho}+\frac{\partial^2h}{\partial z^2}=0.$}. Once the form of $h$ has been specified for a particular solution, the remaining function  $\gamma$ can be obtained by simple quadratures using the equations:
\begin{eqnarray}  \label{gammap1a}
\partial_\rho{\gamma}&=&\rho[(\partial_\rho h)^2-(\partial_z h)^2],~~~~~~~
\partial_z{\gamma}=2\rho(\partial_\rho h)(\partial_z h).
\end{eqnarray}
Also, the function $H$ is the so-called ``dual"of $h$ and it is a solution of the following equation:
\beqs
dH&=&\rho(\partial_{\rho}h dz-\partial_zhd\rho).
\eeqs

Solutions of the pure Einstein-Maxwell theory in five dimensions are simply obtained from the above formulae by taking $\alpha=0$. 

\section{The vacuum KK multi-black hole solution}

\begin{figure}[tbp]
\par
\begin{center}
\includegraphics{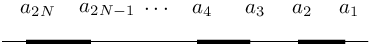} 
\end{center}
\caption{Rod structure of the multi-black hole system.}
\label{KKfig}
\end{figure}

 It has been has been shown in \cite{Stelea:2009ur} that if one starts from the four-dimensional Reissner-Nordstr\"om black hole and use the above solution generating technique one is able to recover the five-dimensional charged KK black hole after setting $a^2=1+c^2$. In this case the harmonic function $h$ is a ``correction" function that depends on the presence of a black hole horizon in the initial seed solution. Then one expects that, in order to construct the five-dimensional generalization of the KK multi-black hole solution, one should make use of the solution previously constructed by Israel and Khan \cite{Khan}, which describes multiple collinear Schwarzschild black holes connected by rods. 
 It turns out that this is indeed the case. In our solution-generating technique, the form of the harmonic function $h$ will now correspond to correction factors applied for each black hole horizon in the four-dimensional Israel-Khan solution. In terms of the ansatz given in (\ref{4dKhan}), the Israel-Khan solution that describes $N$ collinear Schwarzschild black holes is given by:
\beqs
\tilde{f}&=&\prod_{i=1}^N\frac{r_{2i-1}+\zeta_{2i-1}}{r_{2i}+\zeta_{2i}},~~~
e^{2\tilde{\mu}}=\frac{1}{K_0}\left(\frac{4^N}{r_1...r_{2N}}\frac{\prod_{i,j=1}^NY_{2i-1,2j}}{\prod_{i=1}^N\prod_{k>i}Y_{2i,2k}\prod_{i=1}^N\prod_{k>i}Y_{2i-1,2k-1}}\right).
\label{skhan}
\eeqs 
Here we generally denote $\zeta_i=z-a_i$, $r_i=\sqrt{\rho^2+\zeta_i^2}$ while $Y_{ij}=r_ir_j+\zeta_i\zeta_j+\rho^2$ and $K_0$ is an arbitrary constant, fixed in four dimensions by requiring that the asymptotic geometry be flat.\footnote{In what follows we shall keep it unconstrained in the seed solution.} This solution describes then a system of $N$ collinear black holes, having the rods corresponding to the black hole horizons depicted in Figure \ref{KKfig}.

In five dimensions, to describe a configuration of $N$ KK black holes one has to pick the following harmonic function:
\beqs
e^{2h}&=&\prod_{i=1}^N\sqrt{\frac{r_{2i-1}+\zeta_{2i-1}}{r_{2i}+\zeta_{2i}}}.
\label{hcor}
\eeqs
Noting that $e^{2h}=\tilde{f}^{\frac{1}{2}}$, one can actually bypass the integration of (\ref{gammap1a}) by using the scaling symmetry from \cite{Chng:2006gh} to obtain the particularly simple result $\tilde{\mu}=4\gamma$.

Finally, the dual of $h$ turns out to have the particularly simple form\footnote{Up to a constant. In general, the dual of $\frac{1}{2}\ln(r_i+\zeta_i)$ is given by $-\frac{1}{2}(r_i-\zeta_i)$.}:
\beqs
H&=&\frac{1}{4}\sum_{i=1}^N(r_{2i}-r_{2i-1}).
\eeqs

%%%%%%%%%%%%%%%%%%%%%%%%%%%%%%%%%%%%%%%%%%
%%%%%%%%%%%%%%%%%%%%%%%%%%%%%%%%%%%%%%%%%%

To summarize, denoting by  $\Sigma=1+c^2(1-\tilde{f})$, the final five-dimensional solution describing a system of $N$ uncharged KK black holes is given by:
\beqs
ds^2&=&-\tilde{f}dt^2+\frac{1}{\Sigma}\left(d\chi+ac\sum_{i=1}^N(r_{2i}-r_{2i-1})d\varphi\right)^2+\frac{\Sigma}{\tilde{f}}\big[e^{2\tilde{\mu}}(d\rho^2+dz^2)+\rho^2d\varphi^2\big].
\label{finaluncharged}
\eeqs

One can easily check that for $N=2$ this solution corresponds to the uncharged version of the double KK black hole solution previously constructed in \cite{Stelea:2009ur}.

Let us consider now the rod structure of this general solution. For simplicity, let us denote the rod length of each black hole horizon by $2\sigma_i=a_{2i-1}-a_{2i}$. Following the procedure given in \cite{Emparan:2001wk,Harmark:2004rm} one deduces that the rod structure of the general solution is described by $2N$ turning points that divide the $z$-axis into $2N+1$ rods as follows.\footnote{We are writing the vectors in the basis $\{\partial/\partial t, \partial/\partial \chi,\partial/\partial\varphi\}$.} For $z<a_{2N}$ such that all $\zeta_i<0$ one has a semi-infinite rod with direction $l_1=(0,2ac\sum_{i=1}^N\sigma_i,1)$. For $a_{2N}<z<a_{2N-1}$ one has a finite timelike rod with direction $l_2=(1,0,0)$, corresponding to the horizon of the $N$-th black hole. For $a_{2N-1}<z<a_{2N-2}$ one has a spacelike rod with direction $l_3=(0,2ac(\sigma_N-\sum_{i=1}^{N-1}\sigma_i),1)$. More generally, for each black hole horizon $a_{2i}<z<a_{2i-1}$ one has a timelike rod $(1,0,0)$, while in between the black holes (for instance for $a_{2j-1}<z<a_{2j-2}$, which is the rod in between the $(j-1)$-th black hole and the $j$-th black hole) one has a finite spacelike rod $l_{2(N-j)+3}=(0,2ac\left(-\sum_{i=1}^{j-1}\sigma_i+\sum_{i=j}^N\sigma_i\right),1)$. Finally, for $z>a_1$ one has a semi-infinite spacelike rod with direction $l_{2N+1}=(0,-2ac\sum_{i=1}^N\sigma_i,1)$.

Note now that the rod directions of the spacelike rods surrounding the horizons are precisely the rod directions of the multi-Taub-NUT background \cite{Chen:2010zu}. This confirms that the general solution that we derived describes a configuration of black holes sitting at the nuts of the multi-collinearly-centered Taub-NUT background. One can also recover directly the multi-Taub-NUT background by taking the limit in which the black hole horizons disappear. For this, recall that $a^2=1+c^2$ and let us take the limit $c\ra\infty$ and $\sigma_i\ra0$ such that $N_i=c^2\sigma_i$ are kept constant for each $i$. Then, if one denotes $a_{2i-1}=b_i+\sigma_i$ and $a_{2i}=b_i-\sigma_i$ (such that the $i$-th black hole horizon is centered at $b_i$) by expanding to first order in $\sigma_i$ one obtains:
\beqs
\Sigma&=&1+\sum_{i=1}^N\frac{N_i}{\sqrt{\rho^2+(z-b_i)^2}}+{\cal O}(\sigma_i^2),
\eeqs
while
\beqs
acH&=&\sum_{i=1}^N\frac{N_i(z-b_i)}{\sqrt{\rho^2+(z-b_i)^2}}+{\cal O}(\sigma_i^2).
\eeqs
Since in absence of the black holes $\tilde{f}=1$, it is now clear that one recovers as background the multi-collinearly-centered Taub-NUT space with a trivial time direction, as advertised.

We now turn  to the discussion of the conical singularities.  To avoid a conical singularity at the location of a rod with direction $l_i$, the period $\Delta_i$ of the spacelike coordinate $\eta_i$ (such that $l_i=\partial/\partial \eta_i$) must be fixed as:
 \beqs
 \Delta_i=2\pi\lim_{\rho\ra 0}\sqrt{\frac{\rho^2g_{\rho\rho}}{|l_i|^2}},
 \eeqs
 where $g_{\rho\rho}$ is the $\rho\rho$-component of the metric, while $|l_i|^2$ is the norm of $l_i$. More specifically, for the outer axis one has:
 \beqs
 \Delta_1=\Delta_{2N+1}=2\pi\sqrt{\frac{2^{3N}}{K_0}},
 \eeqs
and the conical singularity can be eliminated there by picking $K_0=2^{3N}$. However, in-between the black holes, the expressions for $\Delta_i=2\pi e^{\tilde{\mu}}|_{\rho\ra 0}$ are much more complicated and not informative to list here. It turns out that the conical singularities in-between the black holes cannot be eliminated for any physically reasonable values of the parameters describing the solution. This is, in fact, expected since the multi-black hole solution is static and there are no other forces that could counteract the gravitational attraction between the black holes.

\subsection{Thermodynamic description of KK multi-black holes in presence of conical singularities}

 The geometry describing static configurations of $N$ Schwarzschild black holes has been known for a while \cite{Khan}. Its thermodynamic properties have been investigated in \cite{Costa:2000kf} and more recently in \cite{Herdeiro:2010aq,Herdeiro:2009vd}. As it turns out, even though the geometry has conical singularities in-between the black holes, the multi-black hole system still has a well-defined gravitational action \cite{Gibbons:1979nf}. This means that such black hole solutions with conical singularities might still admit a reasonable well-defined thermodynamic description and it turns out that this is indeed the case. In the usual path-integral description of quantum gravity, the conical singularity manifests itself at the level of the Euclidean action of the system. When a conical singularity is present, the gravitational action gets an extra contribution which is proportional to the conical deficit/excess multiplying the space-time area of the conical singularity's world volume \cite{Herdeiro:2009vd}. More specifically, suppose there is a conical singularity at $\rho=0$ on a finite $z$-interval. To define the conical singularity on a fixed point set of a $U(1)$ isometry with the orbits parameterized by $\eta$, one computes the proper circumference $C$ of these orbits and their proper radius $R$ and one defines
\beqs
\alpha&=&\frac{dC}{dR}|_{R=0}=\lim_{\rho\ra 0}\frac{\partial_{\rho}\sqrt{g_{\eta\eta}}\Delta\eta}{\sqrt{g_{\rho\rho}}}=2\pi k_B,
\eeqs
where in general
\beqs
k_B&=&\lim_{\rho\ra 0}\sqrt{\frac{|l_i|^2}{\rho^2g_{\rho\rho}}}
\eeqs
is the Euclidean surface gravity corresponding to a finite rod with direction $l_i$.
The presence of a conical singularity along a spacelike rod is then expressed by means of the quantity:
\beqs
\delta=2\pi-\alpha=2\pi(1-k_B),
\eeqs
with $\delta>0$, ($\delta<0$) corresponding to a conical deficit, respectively excess. Then, if $Area={\cal A}\beta$ is the space-time area of the conical singularity's world-volume, where $\beta=1/T_H$ is the inverse of the Hawking temperature, the gravitational action receives a contribution proportional to $\delta$ of the form \cite{Herdeiro:2009vd}:
\beqs
 I&=&I_0-\frac{\delta}{8\pi G}{\cal A}\beta.
 \label{fullaction}
 \eeqs 
Here $I_0$ is the action when neglecting the conical singularity. In our case, for vacuum spacetimes with Kaluza-Klein asymptotics, the conical singularity manifests itself as a contribution to the bulk action. Using the Mann-Marolf counterterm to regularize the gravitational action, $I_0$ will correspond to the action computed on the boundary when taking this counterterm into account. In general dimensions, for spaces with Kaluza-Klein asymptotics we follow the general results from \cite{Kleihaus:2009ff}. Assuming one extra-direction (along $\chi$ with the length at infinity $L$) one finds the total mass, gravitational tension, respectively the total action (in absence of conical singularities) to be:
\beqs
M&=&\frac{L\Omega_{D-3}}{16\pi G}\big[(D-3)c_t-c_{\chi}\big],~T=\frac{\Omega_{D-3}}{16\pi G}\big[c_t-(D-3)c_{\chi}\big],~I_0=\frac{\beta L\Omega_{D-3}}{16\pi G}\big[c_t-c_{\chi}\big],
\eeqs
where $\Omega_{D-3}$ is the volume of the $(D-3)$-sphere. Here $c_t$ and $c_{\chi}$ are real constants, which appear in the asymptotic expansions of the metric components $g_{tt}\simeq -1+\frac{c_t}{r^{D-4}}$ and $g_{\chi\chi}\simeq 1+\frac{c_{\chi}}{r^{D-4}}$. One can easily check that one has the relation:
\beqs
I_0&=&\frac{\beta(M+TL)}{D-2}.
\eeqs
For asymptotically flat spaces (in absence of the gravitational tension) this relation reduces to the one found in \cite{Herdeiro:2009vd} in eq. (2.17). One can see that for spaces with Kaluza-Klein asymptotics one has to take into account the effect of the gravitational tension along the extra KK-directions. In the present case, following the discussion presented in \cite{Herdeiro:2009vd} we shall use ${\cal T}_c=-\frac{\delta}{8\pi G}$ and ${\cal A}$ as the thermodynamic variables associated to the conical singularities. In the canonical ensemble in which one keeps the Hawking temperature $T_H$, the \textit{area} ${\cal A}$ and the length $L$ of the KK $\chi$-direction fixed, the free energy becomes:
\beqs
F[T_H,{\cal A},L]=\frac{I}{\beta}={\cal M}-T_HS.
\label{freeenergy}
\eeqs 
Then the entropy $S$, the mass ${\cal M}$, the conical defect tension ${\cal T}_c$ and the gravitational tension $T$ of the system will be given by:
\beqs
S&=&-\frac{\partial F}{\partial T_H}|_{{\cal A},L},~~{\cal M}=F+T_HS,~~{\cal T}_c=\frac{\partial F}{\partial {\cal A}}|_{T_H,L},~~T=\frac{\partial F}{\partial L}|_{T_H,{\cal A}}.
\eeqs
Finally, since the conical singularity manifests itself in the total Euclidean action, it will also lead to a modified first law of black hole thermodynamics:
\beqs
d{\cal M}=T_HdS+TdL+{\cal T}_cd{\cal A}.
\eeqs
It turns out that the mass ${\cal M}$ that enters the first law is related to the conserved mass $M$ by a relation similar to that satisfied in asymptotically flat spaces \cite{Herdeiro:2009vd}:
\beqs
{\cal M}&=&M+{\cal T}_c{\cal A}=M-\frac{\delta}{8\pi G}{\cal A}.
\label{massmod}
\eeqs
This means that the mass ${\cal M}$ is the conserved ADM mass $M$ minus the energy of the strut as seen by a static observer at infinity:
\beqs
E_{int}&=&-{\cal T}_c{\cal A}=\frac{\delta}{8\pi G}{\cal A}.
\eeqs
Let us remark that (\ref{massmod}) is also consistent with the generalized Smarr law (verified for instance in \cite{Stelea:2009ur}). Indeed, replacing the action $I_0$ in (\ref{fullaction}) then the free energy expression (\ref{freeenergy}) leads directly to the following Smarr law:
\beqs
(D-3)M=(D-2)T_HS+TL,
\label{SKK}
\eeqs
in which the entropy for each black hole obeys the usual Bekenstein-Hawking area law. It is remarkable that this law holds even in presence of conical singularities. Finally, let us note that the generalization of these considerations in presence of matter fields can be easily considered.

Further note that the first law can also be written in an equivalent form:
\beqs
dM&=&T_HdS+TdL-{\cal A}d{\cal T}_c,
\label{fl}
\eeqs
when using the conserved mass as computed in the asymptotic region.

\subsection{The Kaluza-Klein double black hole solution}

As an example of the above discussion we shall consider now the thermodynamic properties of the Kaluza-Klein double-black hole solution. In this case one has:
\beqs
\tilde{f}&=&\frac{r_1+\zeta_1}{r_2+\zeta_2}\frac{r_3+\zeta_3}{r_4+\zeta_4},~~~~~~~e^{2\mu}=\frac{16}{K_0}\frac{Y_{12}Y_{14}Y_{23}Y_{34}}{r_1r_2r_3r_4Y_{13}Y_{24}},
\eeqs
and the metric is given in (\ref{finaluncharged}). If one takes $K_0=64$, then according to the general discussion in Section $2$ the rod structure of this solution is as follows: one has four turning points that divide the $z$ axis into five rods. For simplicity, we shall parameterize the turning points as:
\beqs
a_1&=&\frac{R}{2}+\sigma_2,~~~~~a_2=\frac{R}{2}-\sigma_2,~~~~~a_3=-\frac{R}{2}+\sigma_1,~~~~~a_4=-\frac{R}{2}-\sigma_1,
\eeqs
such that the distance between the centres of the two black hole horizons is $R$. Then the rod structure of this solution is given by:
\begin{itemize}
\item For $z<a_4$ one has a semi-infinite spacelike rod, with direction: 
\beqs
l_1&=&(0,2ac(\sigma_1+\sigma_2),1).
\eeqs
\item For $a_4<z<a_3$ one has a finite timelike rod, corresponding to the first black hole horizon, having the direction $l_2=(1,0,0)$. One can compute the surface gravity of this black hole to be:
\beqs
k_{1}&=&\frac{1}{4a}\frac{R+\sigma_1-\sigma_2}{\sigma_1(R+\sigma_1+\sigma_2)}.
\eeqs
\item For $a_3<z<a_2$ one has a finite spacelike rod. Its rod direction is given by $l_3=(0,2ac(\sigma_1-\sigma_2),1)$. The Euclidean surface gravity corresponding to this rod is given by: 
\beqs
k_B&=&\frac{R^2-(\sigma_1-\sigma_2)^2}{R^2-(\sigma_1+\sigma_2)^2}.
\eeqs 
\item For $a_2<z<a_1$ one has a finite timelike rod, corresponding to the second black hole horizon, having the direction $l_4=(1,0,0)$. The surface gravity of the second black hole is:
\beqs
k_{2}&=&\frac{1}{4a}\frac{R+\sigma_2-\sigma_1}{\sigma_2(R+\sigma_1+\sigma_2)}.
\eeqs
\item For $z>a_1$ one has an semi-infinite spacelike rod with direction 
\beqs
l_5=(0,-2ac(\sigma_1+\sigma_2),1).
\eeqs
\end{itemize}

To have the system of two KK black holes in thermodynamic equilibrium, they should have the same temperature. One can satisfy this requirement if one takes the two black holes to have the same mass, that is $\sigma_1=\sigma_2=\sigma$.  Then the Hawking temperature $T_H=\beta^{-1}$ and the area of each black hole horizon are given by:
\beqs
T_H&=&\frac{ R}{8\pi a\sigma(R+2\sigma)}, ~~~~~A_H=\frac{16\pi\sigma^2La(R+2\sigma)}{R}.
\eeqs
Here $L=16\pi ac\sigma$ is the length at infinity of the $\chi$ coordinate, where $a^2=c^2+1$. The total mass and the gravitational tension for the double-black hole solution are computed in the asymptotic region. The more general formulae will be given in the next section, here we shall quote the final results for the two black hole system:
\beqs
M&=&\frac{L\sigma(2+c^2)}{G},~~~~~~~T=\frac{\sigma(1+2c^2)}{G}.
\eeqs
Recall now that the Euclidian surface gravity for the third spacelike rod is $k_B=\frac{R^2}{R^2-4\sigma^2}$ such that $\delta=-\frac{8\pi \sigma^2}{R^2-4\sigma^2}$. The area $\textit{Area}={\cal A}\beta$ of the worldvolume of the conical singularity is easy to compute with the result:
\beqs
 {\cal A}&=&\frac{L(R-2\sigma)}{k_B}.
 \eeqs 
Then the interaction energy between the two KK black holes is:
\beqs
E_{int}&=&\frac{\delta}{8\pi G}{\cal A}=-\frac{L\sigma^2(R-2\sigma)}{R^2G}.
\eeqs
It is now easy to check that the first law (\ref{fl}) is satisfied under independent variations of the parameters $\sigma$, $R$ and $c$ if the entropy of each black hole satisfies the Bekenstein-Hawking relation. Note that the parameters $\sigma$, $R$ and $c$  roughly characterize the mass of each black hole, the distance between them and the asymptotic length of the KK circle. Finally, the Smarr relation (\ref{SKK}) is trivially satisfied.

\section{Multiple charged KK black holes}

The solution-generating technique described in Section $2$ allows us to construct directly the solution describing the superposition of $N$ charged KK black holes; one starts instead with the superposition of $N$ charged Reissner-Nordstr\"om black holes in four dimensions \cite{4dNRN}. In this case, the harmonic function $h$ will be again given by (\ref{hcor}) while the expression for $\gamma=\tilde{\mu}/4$ can be easily read from (\ref{skhan}). However, given the very complicated form of the general solution describing $N$ charged black holes we have chosen to consider here the particular case in which the mass-to-charge ratio is fixed for each black hole. Such a solution can be easily obtained from the uncharged version presented in the previous section by applying a charging technique. 
One particularly simple charging technique has been described in \cite{Gal'tsov:1998yu,Kleihaus:2009ff}. 

Starting from the vacuum solution describing a configuration of $N$ KK black holes one can obtain its dilatonic charged version in the following form:
\beqs
ds^2&=&-\Omega^{-\frac{2}{3\alpha^2+1}}\tilde{f}dt^2+\Omega^{\frac{1}{3\alpha^2+1}}\bigg[\frac{1}{\Sigma}\left(d\chi+ac\sum_{i=1}^N(r_{2i}-r_{2i-1})d\varphi\right)^2+\frac{\Sigma}{\tilde{f}}\big[e^{2\tilde{\mu}}(d\rho^2+dz^2)+\rho^2d\varphi^2\big]\bigg],\nonumber\\
A_t&=&\sqrt{\frac{3}{3\alpha^2+1}}\frac{\tilde{f}U}{\Omega},~~~e^{\phi}=\Omega^{-\frac{3\alpha}{3\alpha^2+1}},~~~~{\rm where} ~~~\Omega=\frac{1-U^2\tilde{f}}{1-U^2},
\eeqs
while $U$ is the parameter of the Harrison transformation, with $0\leq U<1$. For $U=0$ one recovers the vacuum configuration, while $U\ra 1$ corresponds to taking the extremal limit of this charged solution.

Following the analysis performed for the double-black hole case, one is able to compute some of the conserved charges for this multi-black hole configuration. The total mass and the total electric charge are computed in the asymptotic region, which is reached by first performing the following coordinate transformations:
\beqs
\rho&=&r\sin\theta,~~~z=r\cos\theta,
\eeqs
and taking the $r\ra\infty$ limit. Defining now $r_{\infty}=c\sqrt{1+c^2}\sum_{i=1}^N(2\sigma_i)$ the asymptotic length of the $\chi$-circle becomes ${L}=4\pi r_\infty$. Using the counterterm approach \cite{Mann:2005cx,Kleihaus:2009ff} one is now ready to compute the total mass, the gravitational tension and the total electric charge  for this configuration:
\beqs
M&=&\frac{L}{4G}\bigg[\frac{2+U^2}{1-U^2}+c^2+3\alpha^2(c^2+2)\bigg]\sum_{i=1}^N2\sigma_i,~~~T=\frac{1+2c^2}{4G}\sum_{i=1}^N2\sigma_i,~~~
Q=\frac{\sqrt{3}}{3\alpha^2+1}\frac{LU}{4G(1-U^2)}\sum_{i=1}^N2\sigma_i.\nonumber
\eeqs
The dilaton charge can be computed using the asymptotic form of the dilaton field with the result:
\beqs
Q_d=\frac{L}{4G}\frac{3\alpha^2U^2}{(3\alpha^2+1)(1-U^2)}\sum_{i=1}^N2\sigma_i.
\eeqs
One can also compute the so-called Komar mass, check that $2M_K=2M-TL$ and verify that the Smarr relation is satisfied:
\beqs
2M_K&=&3\sum_{i=1}^N\frac{A_{(5)}^ik_{(5)}^i}{8\pi G}+2\Phi Q,
\eeqs
where $\Phi=\Phi_i=\sqrt{\frac{3}{3\alpha^2+1}}U$ is the electric potential of each black hole horizon, while for each black hole one also has:
\beqs
\frac{A_{(5)}^ik_{(5)}^i}{8\pi G}&=&\frac{2\sigma_i L}{4G}.
\eeqs
Here $A_{(5)}^i$ is the horizon area of the $i$-th black hole, while $k_{(5)}^i$ is its surface gravity. As discussed in the previous section, even if conical singularities are present in this system they do not make their appearance into the above Smarr relation. The mass ${\cal M}$ that satisfies the first law of thermodynamics can be related to the conserved mass $M$ by computing the conical singularities in between the black holes, $\delta_i=2\pi(1-k_{Bi})$, and multiplying them to the corresponding space-time areas of the conical singularities' worldvolumes, ${\cal A}_i=Area_i/\beta$, as in (\ref{massmod}). In general, we define $k_{Bi}=\lim_{\rho\ra 0}\sqrt{\frac{|l_i|^2}{\rho^2g_{\rho\rho}}}$ to be the Euclidean surface gravity corresponding to a finite spacelike rod with direction $l_i$.

\subsection{The extremal case}

As we have previously mentioned, the extremal limit of the charged solution is obtained in the limit $U\ra1$ (such that the value of the electric charge is kept finite), which amounts to keeping $M_i=\frac{2\sigma_i}{1-U^2}$ fixed. On the other hand, once $\sigma_i\ra0$ one also has to take the limit $c\ra\infty$ such that $N_i=c^2\sigma_i$ are kept fixed in order to preserve the black holes on the multi-collinearly Taub-NUT background. Gathering up all these results, the extremal solution reduces to:
\beqs
ds^2&=&-\Omega_e^{-\frac{2}{3\alpha^2+1}}dt^2
+\Omega_e^{\frac{1}{3\alpha^2+1}}\bigg[\Sigma_e^{-1}(d\chi+\omega d\varphi)^2+\Sigma_e\left(d\rho^2+dz^2+\rho^2d\varphi^2\right)\bigg],\nonumber\\
A_t&=&\sqrt{\frac{3}{3\alpha^2+1}}\Omega_e^{-1},~~~e^{\phi}=\Omega_e^{-\frac{3\alpha}{3\alpha^2+1}},
\eeqs
where:
\beqs
\Omega_e&=&1+\sum_{i=1}^N\frac{M_i}{\sqrt{\rho^2+(z-b_i)^2}},~~~\Sigma_e=1+\sum_{i=1}^N\frac{N_i}{\sqrt{\rho^2+(z-b_i)^2}},~~~
\omega=\sum_{i=1}^N\frac{N_i(z-b_i)}{\sqrt{\rho^2+(z-b_i)^2}}.\nonumber
\eeqs
This is the general extremal KK multi-black hole solution in the full EMD theory. As expected, for $\alpha=0$ the dilaton vanishes and one recovers the extremal KK multi-black hole solution derived previously in \cite{Ishihara:2006iv}.

\section{Conclusions}

One purpose of this work was to explicitly derive an exact solution describing a superposition of $N$ charged KK black holes in five dimensions. One should note that the solution-generating technique that we used allows us to easily construct the most general solution describing a collinear superposition of $N$ charged KK black holes in five dimensions. For this purpose one should use as the four-dimensional seed solution the general metric constructed previously in \cite{4dNRN}. In this case the harmonic function $h$ is the same as the one used in the vacuum solution in Section $3$. However, due to the complexity of the four-dimensional seed solution and mostly for simplicity reasons, we have chosen to discuss here two particular cases. 

In the third section of this article we focused on the particular case of $N$ neutral KK black holes and studied some of its properties. In particular, we showed that the rod directions of the spacelike rods surrounding the black hole horizons correspond precisely to those of the multi-collinearly-centered Taub-NUT background. We also showed explicitly that in the absence of black holes, one recovers the multi-Taub-NUT background. Even if these exact solutions do exhibit conical singularities in-between the black holes, their gravitational action is still well-defined. We have shown how to properly take into account the effect of the conical singularities and how to relate the physical mass ${\cal M}$ to the conserved ADM mass $M$. It is the physical mass ${\cal M}$ the physical quantity that enters the first law of black hole thermodynamics. We also showed that a Smarr relation is still satisfied, in which the effects of the conical singularities do not show up if one uses $M$ instead of ${\cal M}$. Such a Smarr relation has been previously verified in particular cases in literature.  As an example of the general formalism we developed in Section $3$, we also showed for the particular case of the double KK solution that the first law of thermodynamics as well as the Smarr relation are satisfied when one properly takes into account the effect of the conical singularities. Finally, in Section $4$ we discussed the particular case of $N$ charged KK black holes having fixed mass-to-charge ratio. To generate such a solution from the uncharged version we made use of a charging technique previously discussed in \cite{Gal'tsov:1998yu,Kleihaus:2009ff}. 

By using a counterterm approach we computed the total mass, electric charge and gravitational tension and we showed that the Smarr relation for a configuration of $N$ charged KK black holes in the full EMD theory is satisfied, as expected. Finally, we showed how to obtain the extremal multi-black hole solution of the full EMD theory and recover as a particular case the multi-KK black hole solution of the Einstein-Maxwell theory that was previously derived in \cite{Ishihara:2006iv}. 

As avenues for further research, it would be interesting to investigate the existence of such solutions for more complicated matter fields; for example, for charged Klein-Gordon fields there exist so the called boson star configurations (see for instance \cite{Schunck:2003kk} -\cite{ Dariescu:2002cx} or the more recent review in \cite{Liebling:2012fv}). The collapse of such charged configurations could lead in principle to the formation of charged configurations of multi black holes in $5$ and higher dimensions. Another interesting conjecture has been formulated in \cite{Herdeiro:2014goa} (see also \cite{Brihaye:2014nba}, \cite{Herdeiro:2014ima}) where rotating black holes with nontrivial scalar hair have been found. The rotation of the boson star is necessary in order to be able to add a black hole at its center. One might then inquire if such rotating objects exist in higher dimensional Kaluza-Klein theories as well. This will be the subject of further work.


\begin{thebibliography}{99}

%\cite{Emparan:2001wn}
\bibitem{Emparan:2001wn} R.~Emparan and H.~S.~Reall,
%``A rotating black ring in five dimensions,''
Phys.\ Rev.\ Lett.\ \textbf{88}, 101101 (2002) [arXiv:hep-th/0110260].
%%CITATION = PRLTA,88,101101;%%

%\cite{Emparan:2006mm}
\bibitem{Emparan:2006mm}
  R.~Emparan and H.~S.~Reall,
% ``Black rings,''
  Class.\ Quant.\ Grav.\  {\bf 23}, R169 (2006)
  [arXiv:hep-th/0608012].
  %%CITATION = CQGRD,23,R169;%%
  
%\cite{Emparan:2008eg}
\bibitem{Emparan:2008eg}
  R.~Emparan and H.~S.~Reall,
  %``Black Holes in Higher Dimensions,''
  Living Rev.\ Rel.\  {\bf 11}, 6 (2008)
  [arXiv:0801.3471 [hep-th]].
  %%CITATION = 00222,11,6;%%

%\cite{Obers:2008pj}
\bibitem{Obers:2008pj}
  N.~A.~Obers,
  %``Black Holes in Higher-Dimensional Gravity,''
  Lect.\ Notes Phys.\  {\bf 769}, 211 (2009)
  [arXiv:0802.0519 [hep-th]].
  %%CITATION = LNPHA,769,211;%%  
  
%\cite{Hawking:1971vc}
\bibitem{Hawking:1971vc}
  S.~W.~Hawking,
  %``Black holes in general relativity,''
  Commun.\ Math.\ Phys.\  {\bf 25}, 152 (1972).
  %%CITATION = CMPHA,25,152;%%
  
  %\cite{Chrusciel:1994tr}
\bibitem{Chrusciel:1994tr}
  P.~T.~Chrusciel and R.~M.~Wald,
  %``On the topology of stationary black holes,''
  Class.\ Quant.\ Grav.\  {\bf 11}, L147 (1994)
  [arXiv:gr-qc/9410004].
  %%CITATION = CQGRD,11,L147;%%

%\cite{Jacobson:1994hs}
\bibitem{Jacobson:1994hs}
  T.~Jacobson and S.~Venkataramani,
  %``Topology of event horizons and topological censorship,''
  Class.\ Quant.\ Grav.\  {\bf 12}, 1055 (1995)
  [arXiv:gr-qc/9410023].
  %%CITATION = CQGRD,12,1055;%%

  %\cite{Galloway:1999bp}
\bibitem{Galloway:1999bp}
  G.~J.~Galloway, K.~Schleich, D.~M.~Witt and E.~Woolgar,
  %``Topological censorship and higher genus black holes,''
  Phys.\ Rev.\  D {\bf 60}, 104039 (1999)
  [arXiv:gr-qc/9902061].
  %%CITATION = PHRVA,D60,104039;%%

  %\cite{Galloway:1999br}
\bibitem{Galloway:1999br}
  G.~J.~Galloway, K.~Schleich, D.~Witt and E.~Woolgar,
  %``The AdS/CFT correspondence conjecture and topological censorship,''
  Phys.\ Lett.\  B {\bf 505}, 255 (2001)
  [arXiv:hep-th/9912119].
  %%CITATION = PHLTA,B505,255;%%

%\cite{Galloway:2005mf}
\bibitem{Galloway:2005mf}
  G.~J.~Galloway and R.~Schoen,
  %``A Generalization of Hawking's black hole topology theorem to higher
  %dimensions,''
  Commun.\ Math.\ Phys.\  {\bf 266}, 571 (2006)
  [arXiv:gr-qc/0509107].
  %%CITATION = CMPHA,266,571;%%


%\cite{Myers:1986un}
\bibitem{Myers:1986un} R.~C.~Myers and M.~J.~Perry,
%``Black Holes In Higher Dimensional Space-Times,''
Annals Phys.\ \textbf{172}, 304 (1986). %%CITATION = APNYA,172,304;%%

%\cite{Hollands:2007aj}
\bibitem{Hollands:2007aj}
  S.~Hollands, S.~Yazadjiev,
  %``Uniqueness theorem for 5-dimensional black holes with two axial Killing fields,''
  Commun.\ Math.\ Phys.\  {\bf 283}, 749-768 (2008).
  [arXiv:0707.2775 [gr-qc]].

%\cite{Hollands:2008fm}
\bibitem{Hollands:2008fm}
  S.~Hollands, S.~Yazadjiev,
  %``A Uniqueness theorem for stationary Kaluza-Klein black holes,''
  Commun.\ Math.\ Phys.\  {\bf 302}, 631-674 (2011).
  [arXiv:0812.3036 [gr-qc]].


 %\cite{Dobiasch:1981vh}
\bibitem{Dobiasch:1981vh}
  P.~Dobiasch and D.~Maison,
  % ``Stationary, Spherically Symmetric Solutions Of Jordan's Unified Theory Of
  %Gravity And Electromagnetism,''
  Gen.\ Rel.\ Grav.\  {\bf 14}, 231 (1982).
  %%CITATION = GRGVA,14,231;%%

%\cite{Gibbons:1985ac}
\bibitem{Gibbons:1985ac}
  G.~W.~Gibbons and D.~L.~Wiltshire,
  %``Black Holes In Kaluza-Klein Theory,''
  Annals Phys.\  {\bf 167}, 201 (1986)
  [Erratum-ibid.\  {\bf 176}, 393 (1987)].
  %%CITATION = APNYA,167,201;%%
  
 %\cite{Rasheed:1995zv}
\bibitem{Rasheed:1995zv}
  D.~Rasheed,
  %``The Rotating dyonic black holes of Kaluza-Klein theory,''
  Nucl.\ Phys.\  B {\bf 454}, 379 (1995)
  [arXiv:hep-th/9505038].
  %%CITATION = NUPHA,B454,379;%%

  %\cite{Larsen:1999pp}
\bibitem{Larsen:1999pp}
  F.~Larsen,
  %``Rotating Kaluza-Klein black holes,''
  Nucl.\ Phys.\  B {\bf 575}, 211 (2000)
  [arXiv:hep-th/9909102].
  %%CITATION = NUPHA,B575,211;%%  

%\cite{Sorkin:1983ns}
\bibitem{Sorkin:1983ns}
  R.~D.~Sorkin,
  %``Kaluza-Klein Monopole,''
  Phys.\ Rev.\ Lett.\  {\bf 51}, 87 (1983).
  %%CITATION = PRLTA,51,87;%%

%\cite{Gross:1983hb}
\bibitem{Gross:1983hb}
  D.~J.~Gross and M.~J.~Perry,
 % ``Magnetic Monopoles In Kaluza-Klein Theories,''
  Nucl.\ Phys.\  B {\bf 226}, 29 (1983).
  %%CITATION = NUPHA,B226,29;%%
  
  %\cite{Ishihara:2005dp}
\bibitem{Ishihara:2005dp}
  H.~Ishihara and K.~Matsuno,
 % ``Kaluza-Klein black holes with squashed horizons,''
  Prog.\ Theor.\ Phys.\  {\bf 116}, 417 (2006)
  [arXiv:hep-th/0510094].
  %%CITATION = PTPKA,116,417;%%

  %\cite{Brihaye:2006ws}
\bibitem{Brihaye:2006ws}
  Y.~Brihaye and E.~Radu,
%  ``Kaluza-Klein black holes with squashed horizons and d = 4 superposed
%  monopoles,''
  Phys.\ Lett.\  B {\bf 641}, 212 (2006)
  [arXiv:hep-th/0606228].
  %%CITATION = PHLTA,B641,212;%%

  %\cite{Wang:2006nw}
\bibitem{Wang:2006nw}
  T.~Wang,
 % ``A rotating Kaluza-Klein black hole with squashed horizons,''
  Nucl.\ Phys.\  B {\bf 756}, 86 (2006)
  [arXiv:hep-th/0605048].
  %%CITATION = NUPHA,B756,86;%%

%\cite{Nakagawa:2008rm}
\bibitem{Nakagawa:2008rm}
  T.~Nakagawa, H.~Ishihara, K.~Matsuno and S.~Tomizawa,
%  ``Charged Rotating Kaluza-Klein Black Holes in Five Dimensions,''
  Phys.\ Rev.\  D {\bf 77}, 044040 (2008)
  [arXiv:0801.0164 [hep-th]].
  %%CITATION = PHRVA,D77,044040;%%

 %\cite{Tomizawa:2008hw}
\bibitem{Tomizawa:2008hw}
  S.~Tomizawa, H.~Ishihara, K.~Matsuno and T.~Nakagawa,
  %``Squashed Kerr-Godel Black Holes - Kaluza-Klein Black Holes with Rotations
  %of Black Hole and Universe -,''
  Prog.\ Theor.\ Phys.\  {\bf 121}, 823 (2009)
  [arXiv:0803.3873 [hep-th]].
  %%CITATION = PTPKA,121,823;%%

%\cite{Matsuno:2008fn}
\bibitem{Matsuno:2008fn}
  K.~Matsuno, H.~Ishihara, T.~Nakagawa and S.~Tomizawa,
  %``Rotating Kaluza-Klein Multi-Black Holes with Godel Parameter,''
  Phys.\ Rev.\  D {\bf 78}, 064016 (2008)
  [arXiv:0806.3316 [hep-th]].
  %%CITATION = PHRVA,D78,064016;%%

%\cite{Tomizawa:2008rh}
\bibitem{Tomizawa:2008rh}
  S.~Tomizawa and A.~Ishibashi,
  %``Charged Black Holes in a Rotating Gross-Perry-Sorkin Monopole Background,''
  Class.\ Quant.\ Grav.\  {\bf 25}, 245007 (2008)
  [arXiv:0807.1564 [hep-th]].
  %%CITATION = CQGRD,25,245007;%%

  %\cite{Stelea:2008tt}
\bibitem{Stelea:2008tt}
  C.~Stelea, K.~Schleich and D.~Witt,
 %``On squashed black holes in Godel universes,''
  Phys.\ Rev.\  D {\bf 78}, 124006 (2008)
  [arXiv:0807.4338 [hep-th]].
  %%CITATION = PHRVA,D78,124006;%%

%\cite{Tomizawa:2008qr}
\bibitem{Tomizawa:2008qr}
  S.~Tomizawa, Y.~Yasui and Y.~Morisawa,
  %``Charged Rotating Kaluza-Klein Black Holes Generated by G2(2)
  %Transformation,''
  Class.\ Quant.\ Grav.\  {\bf 26}, 145006 (2009)
  [arXiv:0809.2001 [hep-th]].
  %%CITATION = CQGRD,26,145006;%%

  %\cite{Gal'tsov:2008sh}
\bibitem{Gal'tsov:2008sh}
  D.~V.~Gal'tsov and N.~G.~Scherbluk,
  %``Improved generating technique for D=5 supergravities and squashed
  %Kaluza-Klein Black Holes,''
  Phys.\ Rev.\  D {\bf 79}, 064020 (2009)
  [arXiv:0812.2336 [hep-th]].
  %%CITATION = PHRVA,D79,064020;%%

%\cite{Mizoguchi:2011zj}
\bibitem{Mizoguchi:2011zj} 
  S.~'y.~Mizoguchi and S.~Tomizawa,
  %``New approach to solution generation using SL(2,R)-duality of a dimensionally reduced space in five-dimensional minimal supergravity and new black holes,''
  Phys.\ Rev.\ D {\bf 84}, 104009 (2011)
  [arXiv:1106.3165 [hep-th]].
  %%CITATION = ARXIV:1106.3165;%%
  
%\cite{Myers:1986rx}
\bibitem{Myers:1986rx} R.~C.~Myers,
%``HIGHER DIMENSIONAL BLACK HOLES IN COMPACTIFIED SPACE-TIMES,''
Phys.\ Rev.\ D \textbf{35}, 455 (1987). 
%%CITATION = PHRVA,D35,455;%%

%\cite{Duff:1993ye}
\bibitem{Duff:1993ye} M.~J.~Duff and J.~X.~Lu,
%``Black and super p-branes in diverse dimensions,''
Nucl.\ Phys.\ B \textbf{416}, 301 (1994) [arXiv:hep-th/9306052].
%%CITATION = NUPHA,B416,301;%%

%\cite{Ishihara:2006iv}
\bibitem{Ishihara:2006iv}
  H.~Ishihara, M.~Kimura, K.~Matsuno and S.~Tomizawa,
  %``Kaluza-Klein multi-black holes in five-dimensional Einstein-Maxwell
  %theory,''
  Class.\ Quant.\ Grav.\  {\bf 23}, 6919 (2006)
  [arXiv:hep-th/0605030].
  %%CITATION = CQGRD,23,6919;%%
  
   %\cite{Elvang:2005sa}
\bibitem{Elvang:2005sa}
  H.~Elvang, R.~Emparan, D.~Mateos and H.~S.~Reall,
  %``Supersymmetric 4D rotating black holes from 5D black rings,''
  JHEP {\bf 0508}, 042 (2005)
  [arXiv:hep-th/0504125].
  %%CITATION = JHEPA,0508,042;%

%\cite{Matsuno:2012hf}
\bibitem{Matsuno:2012hf} 
  K.~Matsuno, H.~Ishihara, M.~Kimura and T.~Tatsuoka,
  %``Kaluza-Klein vacuum multi-black holes in five-dimensions,''
  Phys.\ Rev.\ D {\bf 86}, 044036 (2012)
  [arXiv:1206.4818 [hep-th]].
  %%CITATION = ARXIV:1206.4818;%%

%\cite{Chng:2008sr}
\bibitem{Chng:2008sr}
  B.~Chng, R.~Mann, E.~Radu and C.~Stelea,
%``Charging Black Saturn?,''
  JHEP {\bf 0812}, 009 (2008)
  [arXiv:0809.0154 [hep-th]].
  %%CITATION = JHEPA,0812,009;%%
  
%\cite{Stelea:2011jm}
\bibitem{Stelea:2011jm}
  C.~Stelea, C.~Dariescu, M.~-A.~Dariescu,
  %``Static charged double-black rings in five dimensions,''
  Phys.\ Rev.\  {\bf D84}, 044009 (2011).
  [arXiv:1107.3484 [gr-qc]].

%\cite{Tan:2003jz}
\bibitem{Tan:2003jz}
  H.~S.~Tan and E.~Teo,
  %``Multi-black hole solutions in five dimensions,''
  Phys.\ Rev.\  D {\bf 68}, 044021 (2003)
  [arXiv:hep-th/0306044].
  %%CITATION = PHRVA,D68,044021;%%
  
 %\cite{Stelea:2009ur}
\bibitem{Stelea:2009ur}
  C.~Stelea, K.~Schleich and D.~Witt,
  %``Charged Kaluza-Klein double-black holes in five dimensions,''
  Phys.\ Rev.\  D {\bf 83}, 084037 (2011)
  [arXiv:0909.3835 [hep-th]].
  %%CITATION = PHRVA,D83,084037;%%  
  
  %\cite{Costa:2000kf}
\bibitem{Costa:2000kf}
  M.~S.~Costa and M.~J.~Perry,
  %``Interacting black holes,''
  Nucl.\ Phys.\  B {\bf 591}, 469 (2000)
  [arXiv:hep-th/0008106].
  %%CITATION = NUPHA,B591,469;%%
  
  %\cite{Herdeiro:2010aq}
\bibitem{Herdeiro:2010aq}
  C.~Herdeiro, E.~Radu and C.~Rebelo,
  %``Thermodynamical description of stationary, asymptotically flat solutions
  %with conical singularities,''
  Phys.\ Rev.\  D {\bf 81}, 104031 (2010)
  [arXiv:1004.3959 [gr-qc]].
  %%CITATION = PHRVA,D81,104031;%%

%\cite{Herdeiro:2009vd}
\bibitem{Herdeiro:2009vd}
  C.~Herdeiro, B.~Kleihaus, J.~Kunz and E.~Radu,
  %``On the Bekenstein-Hawking area law for black objects with conical
  %singularities,''
  Phys.\ Rev.\  D {\bf 81}, 064013 (2010)
  [arXiv:0912.3386 [gr-qc]].
  %%CITATION = PHRVA,D81,064013;%%
  
 %\cite{Gibbons:1979nf}
\bibitem{Gibbons:1979nf}
  G.~W.~Gibbons and M.~J.~Perry,
  %``NEW GRAVITATIONAL INSTANTONS AND THEIR INTERACTIONS,''
  Phys.\ Rev.\  D {\bf 22}, 313 (1980).
  %%CITATION = PHRVA,D22,313;%%  
  
  \bibitem{Khan} 
W.~Israel and K.~A.~Khan, 
%``Collinear particles and Bondi dipoles in general relativity,'' 
Nuovo Cim.\ {\bf 33} (1964) 331.

  %\cite{Chng:2006gh}
\bibitem{Chng:2006gh}
  B.~Chng, R.~B.~Mann and C.~Stelea,
  %``Accelerating Taub-NUT and Eguchi-Hanson solitons in four dimensions,''
  Phys.\ Rev.\  D {\bf 74}, 084031 (2006)
  [arXiv:gr-qc/0608092].
  %%CITATION = PHRVA,D74,084031;%%
  
  %\cite{Emparan:2001wk}
\bibitem{Emparan:2001wk} R.~Emparan and H.~S.~Reall,
%``Generalized Weyl solutions,''
Phys.\ Rev.\ D \textbf{65}, 084025 (2002) [arXiv:hep-th/0110258].
%%CITATION = PHRVA,D65,084025;%%

%\cite{Harmark:2004rm}
\bibitem{Harmark:2004rm}
  T.~Harmark,
   %``Stationary and axisymmetric solutions of higher-dimensional general
  %relativity,''
  Phys.\ Rev.\  D {\bf 70}, 124002 (2004)
  [arXiv:hep-th/0408141].
  %%CITATION = PHRVA,D70,124002;%%
  
  %\cite{Chen:2010zu}
\bibitem{Chen:2010zu}
  Y.~Chen and E.~Teo,
  %``Rod-structure classification of gravitational instantons with U(1)xU(1)
  %isometry,''
  Nucl.\ Phys.\  B {\bf 838}, 207 (2010)
  [arXiv:1004.2750 [gr-qc]].
  %%CITATION = NUPHA,B838,207;%%
  
  \bibitem{4dNRN}
  N.~Breton, V.~S.~Manko, J.~A.~ Sanchez, 
  %``On the equilibrium of charged masses in general relativity: The electrostatic case", 
  Class.\ Quant.\ Grav. {\bf 15} 3071 (1998)
  
 %\cite{Gal'tsov:1998yu}
\bibitem{Gal'tsov:1998yu}
  D.~V.~Gal'tsov and O.~A.~Rytchkov,
  %``Generating branes via sigma models,''
  Phys.\ Rev.\  D {\bf 58}, 122001 (1998)
  [arXiv:hep-th/9801160].
  %%CITATION = PHRVA,D58,122001;%% 

%\cite{Kleihaus:2009ff}
\bibitem{Kleihaus:2009ff}
  B.~Kleihaus, J.~Kunz, E.~Radu and C.~Stelea,
  %``Harrison transformation and charged black objects in Kaluza-Klein theory,''
  JHEP {\bf 0909}, 025 (2009)
  [arXiv:0905.4716 [hep-th]].
  %%CITATION = JHEPA,0909,025;%%
  
  
 %\cite{Mann:2005cx}
\bibitem{Mann:2005cx}
  R.~B.~Mann and C.~Stelea,
%``On the gravitational energy of the Kaluza Klein monopole,''
  Phys.\ Lett.\  B {\bf 634}, 531 (2006)
  [arXiv:hep-th/0511180].
  %%CITATION = PHLTA,B634,531;%%

 %\cite{Schunck:2003kk}
\bibitem{Schunck:2003kk} 
  F.~E.~Schunck and E.~W.~Mielke,
  %``General relativistic boson stars,''
  Class.\ Quant.\ Grav.\  {\bf 20}, R301 (2003)
  [arXiv:0801.0307 [astro-ph]].
  %%CITATION = ARXIV:0801.0307;%%
  %115 citations counted in INSPIRE as of 19 Oct 2014

%\cite{Dariescu:2010zza}
\bibitem{Dariescu:2010zza} 
  C.~Dariescu and M.~A.~Dariescu,
  %``Boson nebulae charge,''
  Chin.\ Phys.\ Lett.\  {\bf 27}, 011101 (2010).
  %%CITATION = CPLEE,27,011101;%%

%\cite{Dariescu:2003mp}
\bibitem{Dariescu:2003mp} 
  C.~Dariescu and M.~Dariescu,
  %``Transition rates in charged boson nebulae,''
  Phys.\ Lett.\ B {\bf 566}, 19 (2003).
  %%CITATION = PHLTA,B566,19;%%
  %3 citations counted in INSPIRE as of 19 Oct 2014

%\cite{Dariescu:2008zz}
\bibitem{Dariescu:2008zz} 
  C.~Dariescu and M.~A.~Dariescu,
  %``Semiclassical analytic estimation of charged boson nebulae mass and the gravitationally radiated flux,''
  Astropart.\ Phys.\  {\bf 29}, 331 (2008).
  %%CITATION = APHYE,29,331;%%

%\cite{Dariescu:2006tt}
\bibitem{Dariescu:2006tt} 
  M.~A.~Dariescu, C.~Dariescu and G.~Murariu,
  %``Gravitoelectromagnetically induced transitions in charged boson nebulae,''
  Europhys.\ Lett.\  {\bf 74}, 978 (2006).
  %%CITATION = EULEE,74,978;%%
  %1 citations counted in INSPIRE as of 19 Oct 2014

%\cite{Dariescu:2002cx}
\bibitem{Dariescu:2002cx} 
  M.~A.~Dariescu and C.~Dariescu,
  %``First-order perturbative approach to charged boson stars,''
  Phys.\ Lett.\ B {\bf 548}, 24 (2002).
  %%CITATION = PHLTA,B548,24;%%
  %6 citations counted in INSPIRE as of 19 Oct 2014

%\cite{Liebling:2012fv}
\bibitem{Liebling:2012fv} 
  S.~L.~Liebling and C.~Palenzuela,
  %``Dynamical Boson Stars,''
  Living Rev.\ Rel.\  {\bf 15}, 6 (2012)
  [arXiv:1202.5809 [gr-qc]].
  %%CITATION = ARXIV:1202.5809;%%
  %33 citations counted in INSPIRE as of 19 Oct 2014

%\cite{Herdeiro:2014goa}
\bibitem{Herdeiro:2014goa} 
  C.~A.~R.~Herdeiro and E.~Radu,
  %``Kerr black holes with scalar hair,''
  Phys.\ Rev.\ Lett.\  {\bf 112}, 221101 (2014)
  [arXiv:1403.2757 [gr-qc]].
  %%CITATION = ARXIV:1403.2757;%%
  %28 citations counted in INSPIRE as of 19 Oct 2014

%\cite{Brihaye:2014nba}
\bibitem{Brihaye:2014nba} 
  Y.~Brihaye, C.~Herdeiro and E.~Radu,
  %``Myers-Perry black holes with scalar hair and a mass gap,''
  arXiv:1408.5581 [gr-qc].
  %%CITATION = ARXIV:1408.5581;%%
  %2 citations counted in INSPIRE as of 19 Oct 2014

%\cite{Herdeiro:2014ima}
\bibitem{Herdeiro:2014ima} 
  C.~A.~R.~Herdeiro and E.~Radu,
  %``A new spin on black hole hair,''
  arXiv:1405.3696 [gr-qc].
  %%CITATION = ARXIV:1405.3696;%%
  %9 citations counted in INSPIRE as of 19 Oct 2014

 
  
\end{thebibliography}
\end{document}